\documentclass[twocolumn,nofootinbib,preprintnumbers,amsmath,amssymb]{revtex4}

\usepackage{graphicx}
\usepackage{dcolumn}
\usepackage{bm}
\usepackage{amsmath}

\def \be {\begin{equation}}
\def \ee {\end{equation}}

\begin{document}

\title{Black Hole Spindown by Light Bosons}

\author{Andrei Gruzinov}

\affiliation{ CCPP, Physics Department, New York University, 4 Washington Place, New York, NY 10003
}

\begin{abstract}

The saturation mechanism for the fastest-growing instability of massive scalar field in Kerr metric is identified, assuming saturation by cubic or quartic nonlinearities of the field potential. The resulting spindown rate of the black hole is calculated. The (rather involved) saturation scenario is confirmed by numerical simulations.

\end{abstract}

\maketitle

\section{Introduction}

We will assume that bosons with small masses exist and calculate how they spin down black holes. In practice, our results can be used to rule out bosons with certain masses and self-interactions, whenever a spinning black hole is observed. If, on the other hand, an otherwise unexplainable lack of spin of black holes in a certain mass interval is discovered, it can be interpreted as indirect evidence for a light boson \cite{axi, mina}.

\section{Superradiance}

As explained by Zeldovich \cite{zeld}, rotating absorbing body (say a black hole) amplifies waves of frequency $\omega$ and angular momentum projection onto the spin axis $m$, provided $m\Omega > \omega$, where $\Omega$ is the angular velocity of the body. This leads to the so-called superradiant instability of massive boson fields in Kerr metric \cite{det, dol}. 

In Planck units, for the Kerr hole of mass $M$ and dimensionless spin parameter $a$ ($a<1$), the fastest growing mode of the Klein-Gordon field $\phi$ of mass $\mu\ll M^{-1}$ is the 2p state, with the growth rate
\be\label{det}
\gamma _1=\frac{1}{24}aM^8\mu ^9.
\ee

If the number of e-foldings in the black hole lifetime $t_H$ is large ($\gamma_1t_H\gtrsim 100$ amplifies the field from zero-point oscillations to the Planck scale values $\phi \sim M_P$), the exponential growth has to be terminated by the gravitational or $\phi$ radiation. The $\phi$ radiation dominates if the nonlinear scale of the $\phi$  field $f\ll M_P$.  With this assumption, we can study the saturation of the instability by the nonlinearities of the $\phi$ field using the unperturbed Kerr metric and neglecting gravitational radiation.

\section{Nonlinear Saturation of the Instability}\label{sat}

The following (rather involved) saturation scenario has originally been seen in numerical simulations. The numerical results are presented in \S\ref{num}. Here we 
\begin{itemize}

\item (A) give a qualitative picture
\item (B) derive Schrodinger equation for nonrelativistic nonlinear superradiance
\item (C) estimate and then 
\item (D) calculate the mode amplitudes at saturation and the resulting black hole spindown rate. 

\end{itemize}

\subsection{Qualitative Picture}\label{qual}

The crucial feature of our scenario is the simultaneous presence and interaction of the 2p and 3d bound states. As relevant modes are approximately nonrelativistic, they can be analyzed in the Schrodinger equation approximation. Both cubic and quartic self-interactions of the $\phi$ field then give the same $|\Psi |^2\Psi$ nonlinearity in the Schrodinger equation. 

Let $N_1$, $N_2$ be the number of particles in the 2p$(m=1)$ and 3d$(m=2)$ states. The dimensionless nonrelativistic energies (frequencies) of the modes are $\omega _1=-1/8$ and $\omega _2=-1/18$. Let 
\be
\Psi _m\propto N_m^{1/2}e^{-i\omega _mt+im\phi }
\ee
be the corresponding wave functions. 

In the beginning, more precisely, after a few e-foldings, $N_2\ll N_1$. The leading self-interaction is then $|\Psi _1|^2\Psi_1$. This nonlinearity only reshapes the potential well, without affecting the fate of the instability.

The next to leading order interaction is $\Psi _1^2\Psi_2^{*}$. This interaction excites a forced $l=0$ oscillation of the wave function. The frequency of this oscillation is negative, $2\omega _1-\omega _2<0$, so the oscillation is trapped. As $l=0$, the forced oscillation has a finite amplitude at small $r$ and is therefore damped by the black hole. 

The damping reduces the number of particles $N$ , without reducing the angular momentum $L$. Schematically (with arbitrary normalizations):
\begin{align}
& \dot{N}=2\gamma _1N_1-N_1^2N_2 \\
& \dot{L}=2\gamma _1N_1,
\end{align}
where $\gamma _1$ is the Detweiler's growth rate (\ref{det}).
As the forced oscillation amplitude is small, the number of particles and the angular momentum are approximately equal to 
\begin{align}
& N=N_1+N_2,  \label{N}    \\ 
& L=N_1+2N_2.  \label{L}  
\end{align}

It follows that 
\begin{align}
& \dot{N_1}=2\gamma_1N_1-2N_1^2N_2, \\
& \dot{N_2}=N_1^2N_2.
\end{align}
We see that in the presence of $N_1$, the mode $N_2$ becomes linearly unstable with the growth rate $\propto N_1^2$. Formally, the system asymptotes to $N_2\gg N_1$. So, even though we start with $N_2\ll N_1$, the nonlinear interactions will try to reverse the situation. However, as $N_2$ approaches $N_1$, we must also include the interaction $\Psi _2^2\Psi_1^{*}$. This forces the $m=3$ oscillation, which, importantly, is unbound:
\be
2\omega _2-\omega_1>0
\ee
The $m=3$ oscillation carries particles and angular momentum to infinity. Again schematically, treating eqs.(\ref{N}, \ref{L}) as the definitions of $N$ and $L$, we have
\begin{align}
& \dot{N}=2\gamma _1N_1-N_1^2N_2 -N_1N_2^2\\
& \dot{L}=2\gamma _1N_1-3N_1N_2^2,
\end{align}
giving 
\begin{align}
& \dot{N_1}=2\gamma_1N_1-2N_1^2N_2+N_1N_2^2,  \label{sk1} \\
& \dot{N_2}=N_1^2N_2-2N_1N_2^2.               \label{sk2}
\end{align}

Equations (\ref{sk1}, \ref{sk2}) have a single equilibrium point $N_1=2N_2=(8\gamma _1/3)^{1/2}$, which is stable for all $\gamma_1$. At equilibrium, particles and angular momentum are extracted from the black hole. Then some particles are returned to the black hole and some particles, carrying all the extracted angular momentum, are radiated to infinity.

\subsection{Schrodinger equation}

To proceed, we need Schrodinger equation describing nonrelativistic nonlinear superradiance. Consider Klein-Gordon equation with cubic and quartic nonlinearities in the field potential:
\be
V=\frac{1}{2}\mu ^2\phi^2+\frac{\lambda _3}{3!}\frac{\mu ^2}{f}\phi^3+\frac{\lambda _4}{4!}\frac{\mu ^2}{f^2}\phi^4,
\ee
where $f$ is the nonlinear scale of the field $\phi$. The coefficients $\lambda_3$, $\lambda_4$ are dimensionless. For example, $\lambda _3=0$, $\lambda _4=-1$ give the first expansion terms of the axion potential.

For $M\mu \ll 1$, the growing mode and all other modes participating in the nonlinear saturation of the instability are nonrelativistic. The nonlinear Klein-Gordon equation then reduces to the nonlinear Schrodinger equation. Namely, define $\Psi$ by
\be
\phi = \mu^{-1/2}\Psi e^{-i\mu t}+\mu^{-1/2}\Psi ^*e^{i\mu t}.
\ee
Then, with Newtonian gravitational field, we have
\be\label{sch0}
i\dot{\Psi}=-\frac{1}{2\mu }\nabla ^2 \Psi-\frac{M\mu }{r}\Psi-\lambda |\Psi |^2\Psi .
\ee
Here
\be
\lambda \equiv \frac{\lambda_3^2/3-\lambda_4/4}{f^2}=\pm \frac{1}{4f^2},
\ee
with the last equality redefining the nonlinear scale $f$ (we do not consider the degenerate case $\lambda_3^2/3=\lambda_4/4$). In the axion case, $\lambda >0$, meaning that the self-interaction is attractive.

As it is, eq.(\ref{sch0}) does not have any growing or decaying modes, because the near-horizon effects disappear in the nonrelativistic approximation. But, as will become clear from what follows, all we need for our calculation is: (i) correct growth rates for the 2p mode with $m=1$, (ii) correct decay rates for the $l=0$ modes, (iii) negligibly small growth/decay rates for $l>1$. This can be achieved by adding a rotating absorber to eq.(\ref{sch0}):
\begin{multline}\label{se}
i\dot{\Psi}=-\frac{1}{2\mu }\nabla ^2 \Psi-\frac{M\mu }{r}\Psi-\lambda |\Psi |^2\Psi \\
+\sigma \theta (b-r)(\frac{\Omega }{\mu}\partial _\phi-i)\Psi. 
\end{multline}

Here $\sigma$ is the decay rate, $b$ is the radius of the absorber, $\Omega$ is the angular velocity. So far these parameters are arbitrary. However, only with 
\be
b=\left(\frac{5}{3}\right)^{1/2}M, ~~~ \Omega=\frac{a}{2r_+},~~~ \sigma = 6\left(\frac{3}{5}\right)^{3/2}\frac{r_+}{M^2},
\ee
the first-order perturbation theory growth rates of eq.(\ref{se}),
\begin{multline}
\gamma _{nl} = \frac{2^{2l+2}(n+l)!}{n^{2l+4}(n-l-1)!(2l+1)!^2(2l+3)} \\
\sigma \left( \frac{m\Omega}{\mu}-1\right)(Mb\mu^2)^{2l+3},
\end{multline}
coincide with those given by Detweiler (for all $l=0$ and for 2p in the limit of small $\mu$; note that our $n$ is the standard principle quantum number, with $n>l$).

\subsection{Estimates}

To convert the schematics of \S\ref{qual} into estimates, we note that the Schrodinger equation (\ref{se}) gives the length scales of the modes
\be
r\sim (M\mu )^{-1}\mu ^{-1},
\ee
frequencies
\be
\omega \sim \mu ^{-1}r^{-2}\sim (M\mu)^2\mu,
\ee
amplitudes, in terms of the particle numbers $N$,
\be
\Psi \sim N^{1/2}r^{-3/2},
\ee
and nonlinearly induced amplitudes
\be
\Psi _{\rm ind} \sim \omega ^{-1}\lambda \Psi ^3\sim \left(\frac{\mu}{f}\right)^{2}(M\mu)^{5/2}\mu^{3/2}N^{3/2}.
\ee

The induced $m=0$ mode is damped by the absorber (the black hole) at the rate
\be
\dot{N_0}\sim -\sigma b^3\Psi _{\rm ind}^2,
\ee
giving
\be
\dot{N_0}=-\gamma _0N_1^2N_2,
\ee
\be
\gamma _0=\tilde{\gamma _0}\frac{r_+}{M}(M\mu)^{7}\left(\frac{\mu}{f}\right)^{4}\mu,
\ee
where $\tilde{\gamma _0}$ is a dimensionless coefficient to be calculated in the next section.

The induced $m=3$ mode is radiated at the rate 
\be
\dot{N_3}\sim -vr^2\Psi _{\rm ind}^2,
\ee
where the velocity is
\be
v\sim (\mu ^{-1}\omega )^{1/2}\sim M\mu,
\ee
giving
\be
\dot{N_3}=-\gamma _3N_1N_2^2,
\ee
\be
\gamma _3=\tilde{\gamma _3}(M\mu)^{4}\left(\frac{\mu}{f}\right)^{4}\mu,
\ee
where $\tilde{\gamma _3}$ is a dimensionless coefficient to be calculated in the next section.

We thus have
\begin{align}
& \dot{N}=2\gamma _1N_1-\gamma_0N_1^2N_2 -\gamma_3N_1N_2^2\\
& \dot{L}=2\gamma _1N_1-3\gamma_3N_1N_2^2,
\end{align}
giving
\begin{align}
& \dot{N_1}=2\gamma_1N_1-2\gamma_0N_1^2N_2+\gamma_3N_1N_2^2,   \\
& \dot{N_2}=\gamma_0N_1^2N_2-2\gamma_3N_1N_2^2.              
\end{align}

The saturated particle numbers are
\be
N_1=\left(\frac{8\gamma_1\gamma _3}{3\gamma_0^2}\right)^{1/2}, ~~~N_2=\left(\frac{2\gamma_1}{3\gamma_3}\right)^{1/2}.
\ee
The resulting torque on the black hole is $K=2\gamma_1N_1$, or
\be
K=\frac{1}{36}\frac{\tilde{\gamma _3}^{1/2}}{\tilde{\gamma _0}}a^{3/2}\frac{M}{r_+}(M\mu)^7\frac{f^2}{\mu^2}\mu
\ee

\subsection{Final Results and Calculations}
We start by listing the final results. Then we outline the calculations. 

\subsubsection{Final Results}
We will show that
\be
\tilde{\gamma _0}\approx 3.4\times 10^{-6},~~~\tilde{\gamma _3}\approx 4.5\times 10^{-8}.
\ee
Then the torque on the black hole is
\be
K\approx 1.7a^{3/2}\frac{M}{r_+}(M\mu)^7\frac{f^2}{\mu^2}\mu
\ee
We must also make sure that the instability saturates at $\phi <f$, otherwise our calculation would be meaningless. And, indeed, we get for the maximal values of the $m=1$ and the $m=2$ modes
\be
\frac{\phi _{1{\rm max}}}{f}\approx 0.4a^{1/4}\left(\frac{M}{r_+}\right)^{1/2}(M\mu ),
\ee
\be
\frac{\phi _{2{\rm max}}}{f}\approx a^{1/4}(M\mu )^{5/2}.
\ee

\subsubsection{Calculations}

With $M=\mu =1$, the wave functions of the 2p$(m=1)$ and 3d$(m=2)$ modes are
\be
\Psi_1=N_1^{1/2}R_{21}Y_{11}e^{-i\omega _1t},~~\Psi_2=N_2^{1/2}R_{32}Y_{22}e^{-i\omega _2t}
\ee
where
\be
\omega _m=-\frac{1}{2(m+1)^2},
\ee
\be
R_{21}=\frac{1}{2\sqrt{6}}re^{-r/2},~~~ R_{32}=\frac{4}{81\sqrt{30}}r^2e^{-r/3}.
\ee

{\bf The $\Psi_1^2\Psi _2^{*}$ forced oscillation:} We first calculate the product of spherical harmonics:
\be
Y_{11}^2Y_{22}^{*}=\frac{1}{2\pi}\left(\frac{15}{2}\right)^{1/2}\left(\frac{1}{35}Y_{40}+\frac{2}{7\sqrt{5}}Y_{20}+\frac{1}{5}Y_{00} \right).
\ee
Only the $l=0$ forced oscillation will penetrate to the absorber and damp the particle number. The $l=0$ forced oscillation is
\be
\Psi _{\rm ind}=\Psi(r)Y_{00}e^{-i\omega _{\rm ind}t},
\ee
\be
\omega _{\rm ind}=2\omega _1-\omega _2=-\frac{7}{36}.
\ee
The (first-order perturbation theory) damping rate is
\be
\dot{N_0}=-\frac{2}{3}\sigma b^3|\Psi(0)|^2,
\ee
and $\Psi(r)$ satisfies
\be
\Psi''+\frac{2}{r}\Psi'+\frac{2}{r}\Psi-\frac{7}{18}\Psi=Ar^4e^{-\frac{4}{3}r},
\ee
\be
A=-\frac{1}{4\pi \cdot 5\cdot 3^5}\lambda N_1N_2^{1/2}.
\ee
Numerical integration of 
\be
F''+\frac{2}{r}F'+\frac{2}{r}F-\frac{7}{18}F=r^4e^{-\frac{4}{3}r},
\ee
gives (for the solution which asymptotes to zero at infinity) $|F(0)|=C_0$, where
\be
C_0\approx 56.
\ee
We have
\be
|\Psi(0)|=\frac{C_0}{4\pi \cdot 5\cdot 3^5}\lambda N_1N_2^{1/2},
\ee
and, finally,
\be
\dot{N_0}=-\gamma_0N_1^2N_2,~~~\tilde{\gamma_0}\approx 3.4\times 10^{-6}.
\ee

{\bf The $\Psi_2^2\Psi _1^{*}$ forced oscillation:} The product of spherical harmonics:
\be
Y_{22}^2Y_{11}^{*}=\frac{1}{2\pi}\left(\frac{5}{42}\right)^{1/2}\left(2Y_{33}+\frac{1}{\sqrt{11}}Y_{53}\right).
\ee
Both the $l=3$ and the $l=5$ forced oscillations radiate the particle number and angular momentum to infinity. The $l$ forced oscillation is
\be
\Psi _{\rm ind}=\Psi(r)Y_{l3}e^{-i\omega _{\rm ind}t},
\ee
\be
\omega _{\rm ind}=2\omega _2-\omega _1=\frac{1}{72}.
\ee
The radiation rate is
\be
\dot{N_3}=-k\sum _{l=3,5}(r|\Psi(r)|)^2,~~~r\rightarrow \infty,
\ee
\be
k=\sqrt{2\omega _{\rm ind}}=\frac{1}{6}.
\ee
$\Psi(r)$ satisfies
\be
\Psi''+\frac{2}{r}\Psi'-\frac{l(l+1)}{r^2}\Psi+\frac{2}{r}\Psi+k^2\Psi=A_lr^5e^{-\frac{7}{6}r},
\ee
\be
A_3=-\frac{4}{\pi \cdot 3^{10}\sqrt{35}}\lambda N_1^{1/2}N_2,
\ee\be
A_5=-\frac{2}{\pi \cdot 3^{10}\sqrt{385}}\lambda N_1^{1/2}N_2.
\ee
Numerical integration of 
\be
F''+\frac{2}{r}F'-\frac{l(l+1)}{r^2}F+\frac{2}{r}F+k^2F=r^5e^{-\frac{7}{6}r},
\ee
gives (for the solution which asymptotes to an outgoing mode at infinity, for large $r$) $r|F(r)|=C_l$, where
\be
C_3\approx 570,~~~C_5\approx 80.
\ee
We have, for large $r$,
\be
r|\Psi _3(r)|=\frac{4C_3}{\pi \cdot 3^{10}\sqrt{35}}\lambda N_1^{1/2}N_2,
\ee\be
r|\Psi _5(r)|=\frac{2C_5}{\pi \cdot 3^{10}\sqrt{385}}\lambda N_1^{1/2}N_2,
\ee
and, finally,
\be
\dot{N_3}=-\gamma_3N_1N_2^2,~~~\tilde{\gamma_3}\approx 4.5\times 10^{-8}.
\ee

\begin{figure}[bth]
  \centering
  \includegraphics[width=0.48\textwidth]{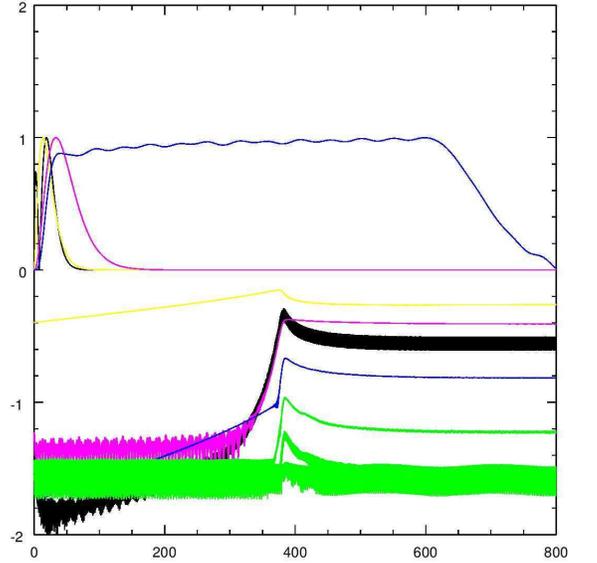}
\caption{Upper curves: (arbitrarily normalized) $|\phi _m(r)|r^{1/2}$ vs. $r$ at the end of the simulation for $m=0,1,2,3$ harmonics (black, yellow, magenta, blue). Lower curves: $\log _{1000}|\phi _m|_{\rm max}$ as a function of (arbitrarily normalized) time for the entire simulation (same color coding for $m=0,1,2,3$, and green for $m=4,5,6$.} 
\end{figure}

\section{Numerical Simulations}\label{num}

The figures show simulation results for the ``heavy Zeldovich cylinder'' -- two dimensional self-interacting scalar field in the presence of a ``gravitating'' rotating absorber, described by 
\begin{multline}
\ddot{\phi}=\nabla ^2 \phi-\mu ^2(1-\frac{2r_g}{r})(\phi -\frac{\phi ^3}{6})-\\ 
\sigma \theta (r_g-r)(\partial _t+\Omega \partial _\theta )\phi,
\end{multline}
with
\be\label{param}
r_g=1,~~\Omega =0.5,~~\mu=0.4,~~\sigma=3.
\ee
The calculation is done by decomposing
\be
\phi(t,r,\theta)=\sum _{m=-m_{\rm max}}^{m_{\rm max}}\phi_m(t,r)e^{im\theta}, ~~~\phi_{-m}=\phi _m^{*},
\ee
and solving the resulting set of $m_{\rm max}+1$ coupled (1+1)d equations. As unbound modes do get excited, we put a ``damping buffer'' into the last quarter of the simulated $r$ domain.

So long as $m_{\rm max}\geq 3$ the number of modes does not matter. This is the way it should be theoretically, and this what the figures show -- other modes are not efficiently excited.

We have also simulated the cases of cubic potential and repulsive quartic potential. As expected (\S \ref{qual}), we got similar results, but with one interesting exception. For quartic repulsion, with not too small $\mu $, like what's listed in eq.(\ref{param}), the saturation scenario is entirely different. Repulsion reduces the mode-mode interaction so much that the $m=2$ mode never grows to a significant level. Instead the most unstable $m=1$ mode grows asymptotically to the marginally bound state:
\be
\phi_1 ''+\frac{1}{r}\phi_1 '-\frac{1}{r^2}\phi_1-\mu ^2(1-\frac{2r_g}{r})(\phi_1 +\frac{\phi_1 ^3}{2})=-\mu ^2\phi_1 .
\ee
As this state has an infinite number of particles ($\phi_1 \propto r^{-1/2}$ at large $r$), the growth of the number of particles lasts forever, while the maximal value of $\phi$ saturates.

We see that numerics does confirm the theoretical saturation scenario. There is no ``bose-nova'' . Rather a steady state is formed, with boson radiation from the black hole \cite{hir}.

\section{Conclusion}

A physically clear, even if somewhat intricate, scenario for the nonlinear saturation of superradiance is proposed: (i) black hole (BH) pumps 2p, (ii) 2p pumps 3d, (iii) 2p and 3d pump unbound f. In more details:
\begin{itemize}
\item BH pumps particles, $N$, and angular momentum, $L$, into 2p
\item interaction (2p)$^2$(3d) pumps bound s
\item s pumps $N$ but not $L$ back into BH, meaning that 2p pumps 3d
\item interaction (3d)$^2$(2p) pumps unbound f
\end{itemize}
The resulting torque on the black hole, for $M\mu \ll 1$, is 
\be
K\sim a^{3/2}(M\mu)^7\frac{f^2}{\mu^2}\mu.
\ee

\begin{figure}[b]
  \centering
  \includegraphics[width=0.48\textwidth]{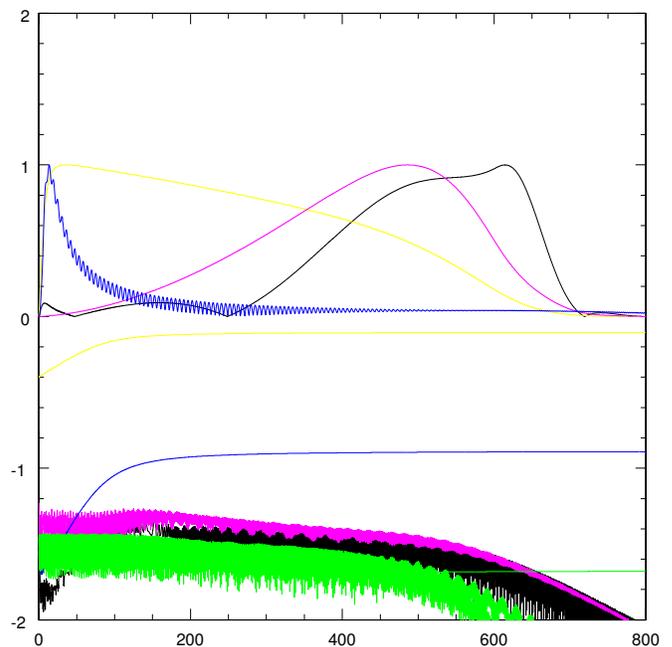}
\caption{Same as fig.1, but for repulsive self-interaction.} 
\end{figure}

\thanks

I thank Sergei Dubovsky for many discussions. I thank Mina Arvanitaki and Sergei Dubovsky for explaining to me the potential uses of these results.


\begin{thebibliography}{99}

\bibitem{axi}
A. Arvanitaki, S. Dimopoulos, S. Dubovsky, N. Kaloper, J. March-Russell,  Phys. Rev. D, 81, 123530 (2010)

\bibitem{mina}
A. Arvanitaki, M. Baryakhtar,X. Huang, Phys. Rev. D, 91, 084011 (2015)

\bibitem{zeld}
Ya. B. Zeldovich, JETP Letters, 14, 180 (1971)

\bibitem{det}
S. Detweiler, Phys. Rev. D, 22, 2323 (1980)

\bibitem{dol}
S. R. Dolan, Phys. Rev. D, 76, 084001 (2007)

\bibitem{hir}
H. Yoshino, H. Kodama, C.Q. Gra., 32, 214001 (2015)

\end{thebibliography}
\end{document}